\documentclass[%
 aip,
 jcp,
 amsmath,amssymb,
 reprint,%
]{revtex4-2}

\usepackage{amsmath}
\usepackage{amsfonts}
\usepackage{amsthm}
\usepackage{graphicx}
\usepackage{dcolumn}
\usepackage{bm}

\usepackage[utf8]{inputenc}
\usepackage[T1]{fontenc}
\usepackage{mathptmx}
\usepackage{etoolbox}
\usepackage{comment}
\usepackage{amsmath}
\usepackage{physics}
\usepackage{siunitx}
\usepackage{float}

\begin{document}


\title{Noise Reduction of Stochastic Density Functional Theory for Metals}
\author{Jake P. Vu}
\email{vuj@purdue.edu}
\affiliation{Department of Chemistry, Purdue University, West Lafayette, Indiana 47906, USA}

\author{Ming Chen}%
\email{chen4116@purdue.edu}
\affiliation{Department of Chemistry, Purdue University, West Lafayette, Indiana 47906, USA}%

\date{\today}

\begin{abstract}
Density Functional Theory (DFT) has become a cornerstone in the modeling  
of metals. However, accurately simulating metals, particularly under 
extreme conditions, presents two significant challenges. First, simulating complex 
metallic systems at low electron temperatures is difficult due to their highly 
delocalized density matrix. Second, modeling metallic warm-dense materials 
at very high electron temperatures is challenging because it requires the 
computation of a large number of partially occupied orbitals. 
This study demonstrates that both challenges can be effectively addressed using the latest advances in linear-scaling stochastic DFT methodologies. 
Despite the inherent introduction of noise into all computed properties by stochastic DFT, 
this research evaluates the efficacy of various noise reduction techniques under different 
thermal conditions. Our observations indicate that the effectiveness of noise reduction 
strategies varies significantly with the electron temperature. Furthermore, we provide 
evidence that the computational cost of stochastic DFT methods scales linearly with 
system size for metal systems, regardless of the electron temperature regime.
\end{abstract}

\maketitle

\section{\label{sec:level1}Introduction}
Metals are ubiquitous in daily life, industry, and academic research, due to their
unique mechanical, magnetic, and electronic properties
\cite{bhaduri2018mechanical,MCHENRY20141881,EHRENREICH1976149}. Theoretical modeling of 
metal electronic structures is crucial for understanding their ground and excited state 
properties \cite{aarons2016perspective,Matthew2023plasmon}. Density functional theory
(DFT) is a widely used approach in materials science and condensed matter physics 
to accurately model the ground-state properties of metals at reasonable computational
costs. \cite{aarons2016perspective,PhysRevB.27.5169,doi:10.1146/annurev.physchem.53.100301.131630}
Recent studies have focused on complex metal systems such as metal nanostructures, 
high-entropy alloys, and metal-support interfaces, which exhibit intriguing  
properties and have significant applications 
\cite{doi:10.1021/acs.accounts.9b00224,Gao2022,doi:10.1021/cm402131n}. These materials 
often require large supercells in theoretical modeling, sometimes containing more than 
1000 atoms 
\cite{doi:10.1021/acsomega.3c00502,PhysRevB.93.155302,doi:10.1021/acs.nanolett.6b04799}. 
While DFT has been effective for studying bulk metal systems, the computational cost of 
conventional DFT scales as $O(N^{3}_{e})$, where $N_{e}$ represents the number of electrons. 
This makes the application of conventional DFT methods to complex metal systems 
challenging.

Another significant challenge arises in modeling metallic systems at high electron 
temperatures, especially in the context of warm-dense materials relevant to fusion 
energy \cite{koenig2005progress}. Accurate modeling of these materials requires the 
description of electronic structure at extremely high electron temperatures 
\cite{smith2018warming}. Conventional DFT methods face difficulties in modeling warm dense 
materials due to the requirement of a large number of partially occupied orbitals 
\cite{blanchet2020requirements}. Therefore, the development of DFT methodologies that can 
minimize computational expenses is essential for modeling metals across various system 
sizes and under different conditions.

Various linear-scaling DFT methods have been developed to address the computational challenges 
associated with modeling complex materials \cite{PhysRevB.53.12713, prentice2020onetep, kohn1996density, 
wesolowski2013recent, artacho1999linear, hernandez1996linear, henderson2006embedding, huang2008advances, 
Soler_2002, Krishtal_2015}. These methods generally employ one of two strategies: exploiting the 
``nearsightedness'' of the electronic structure, or decomposing the system into smaller, manageable 
subsystems \cite{kohn1996density, PhysRevLett.66.1438, huang2008advances, 
https://doi.org/10.1002/wcms.1175, wesolowski2015frozen, Krishtal_2015}.
The concept of nearsightedness in electronic structure implies that the one-body density 
matrix, $\rho(\mathbf{r}, \mathbf{r}')$, decays exponentially with the distance between points 
$\mathbf{r}$ and $\mathbf{r}'$ \cite{PhysRevLett.76.3168, PhysRevB.62.12573}. 
This property allows for the truncation of the density matrix, facilitating the use of sparse matrix 
techniques to achieve linear-scaling computational efficiency \cite{Baer1997, challacombe1999simplified}. 
Such an approach is particularly effective for systems with a large band gap or metallic systems at high 
electron temperatures \cite{Benzi2012, Baer1997}. However, challenges arise when dealing with metals at 
low electron temperatures due to the slow decay of $\rho(\mathbf{r}, \mathbf{r}')$, making the 
application of this principle less straightforward \cite{IsmailBeigi1999, ismail1999locality}.
Alternatively, some methods focus on the subsystem decomposition approach, which relies on embedding 
theory to account for interactions between a subsystem and its environment \cite{PhysRevB.53.12713}. 
This technique has shown success with non-covalently bonded systems, such as molecular clusters 
\cite{huang2008advances}, but faces significant challenges when applied to inorganic materials with 
covalent bonds \cite{Sun2016}.
Notwithstanding, certain linear-scaling DFT methods have demonstrated the ability to efficiently model 
metals with high electron temperatures by focusing on a localized density matrix. Nonetheless, these 
methods often struggle to access high-energy orbital information, which remains a limitation 
\cite{prentice2020onetep, skylaris2005introducing, white2020fast}.

Stochastic DFT (sDFT) represents a significant advancement in linear-scaling DFT methodologies, uniquely 
addressing challenges posed by systems characterized by small or negligible fundamental gaps 
\cite{baer2013self}. Unlike traditional DFT approaches that depend on the Kohn-Sham (KS) orbitals, 
sDFT calculates ground-state properties—including electron density, ground-
state energy, and forces on nuclei—through the statistical averaging over a set of stochastic orbitals. 
Research into the application of sDFT on semiconductor materials has revealed a particularly compelling 
advantage: the number of stochastic orbitals required for accurate property calculation does not scale 
with the size of the system. This characteristic enables sDFT to achieve linear or even sublinear scaling 
efficiency for computations of electron density, energy per particle, and nuclear forces. 
The independence from explicit KS orbitals allows sDFT to effectively model systems at high electron 
temperatures, such as warm-dense materials (WDM), with enhanced computational efficiency 
\cite{PhysRevB.97.115207,hadad2024stochastic,white2020fast,PhysRevB.106.125132}.

Inherent to sDFT, stochastic noise affects all calculated properties, presenting a significant 
computational challenge. To reduce this noise by an order of magnitude, the number of stochastic orbitals 
must be increased by two orders of magnitude, leading to a substantial rise in computational demand 
\cite{10.1063/1.5114984}. To mitigate this issue, a variety of noise reduction techniques have been 
developed, each leveraging different approaches to enhance computational efficiency without compromising 
accuracy.
Among these, ``overlapped embedded-fragmented sDFT'' (o-efsDFT) 
which is based on 
real-space fragmentation technique (oef-sDFT) \cite{10.1063/1.5064472}, ``energy-window sDFT'' (ew-sDFT) 
which is an energy-space 
fragmentation method \cite{10.1063/1.5114984}, and ``energy window embedded-fragmented sDFT'' 
(ew-efsDFT) which uses a hybrid strategy 
\cite{10.1063/5.0044163} have shown promise. These methods facilitate the study of semiconductors, 
including those with minimal band gaps, by effectively managing stochastic noise and computational 
workload. Despite their success with semiconductors, the application of these noise reduction techniques 
to metals, particularly at varying electron temperatures, remains an area with limited exploration 
\cite{hadad2024stochastic}.


In this study, we undertake a comprehensive benchmarking of noise reduction techniques in stochastic 
Density Functional Theory (sDFT), aimed at efficiently modeling metals subjected to both low and high 
electron temperatures. Our paper is structured to facilitate a clear understanding of these techniques 
and their efficacy.
We begin by offering a concise introduction to sDFT, alongside a detailed overview of the various noise 
reduction strategies that have been developed to date. This sets the foundation for our subsequent 
analysis.
Following the introduction, we delve into an empirical evaluation of these noise reduction techniques, 
using bulk aluminum as our test system. This evaluation encompasses simulations at both room temperature 
and elevated temperatures, providing insights into the performance of these methods across a range of 
thermal conditions.
Additionally, we extend our analysis to compare the computational costs associated with the ew-efsDFT 
method across different system sizes and temperatures. This comparison aims to elucidate the scalability 
and efficiency of ew-efsDFT, offering valuable perspectives on its practical application in materials 
science research.
Through this structured approach, our study aims to illuminate the capabilities and limitations of noise 
reduction techniques in sDFT, contributing to the ongoing optimization of computational methodologies for 
the modeling of metal systems under diverse thermal conditions.

\section{Stochastic Density Functional Theory}
We initiate our discussion with a consideration of a supercell of volume $V$, encompassing an electron density $\rho(\mathbf{r})$ that is discretized over a real-space grid consisting of $N_{G}$ grid points. For the scenario of spin-unpolarized systems, the electron density is expressed as: 
\begin{equation}
    \rho(\textbf{r}) = 2\Sigma_{i}f(\varepsilon_i)\langle\textbf{r}|\psi_{i}\rangle\langle\psi_{i}|\textbf{r}\rangle
    \label{eq:rho}
\end{equation}
where $\psi_{i}(\mathbf{r}) = \langle\mathbf{r}|\psi_{i}\rangle$ represents the Kohn-Sham (KS) 
orbitals, and $\varepsilon_i$ denotes the corresponding KS orbital energies, according to the 
Kohn-Sham formulation of DFT. The function $f(x)$, serving as a smearing function to accommodate 
the occupancy of states, is crucial for DFT calculations in metallic systems. Specifically, we 
employ the Fermi-Dirac distribution:
\begin{equation}
    f(x;\mu,\beta)=\frac{1}{1+e^{\beta(x-\mu)}}
    \label{eq:fermi}
\end{equation}
with $\mu$ symbolizing the chemical potential, and $\beta = 1/k_{\mathrm{B}}T$ representing the 
inverse temperature factor, where $k_{\mathrm{B}}$ is the Boltzmann constant and $T$ denotes the 
electron temperature. 
It is pertinent to highlight that alternative smearing functions, such as the error function, are equally viable for implementation within the scope of sDFT.
The Kohn-Sham Hamiltonian, $\hat{h}_{\mathrm{KS}}$, is defined as:
\begin{equation}
\hat{h}_{\mathrm{KS}} = \hat{t} + \hat{v}_{\mathrm{loc}} + \hat{v}_{\mathrm{nl}} + 
\hat{v}_{\mathrm{H}} + \hat{v}_{\mathrm{XC}},
\end{equation}
where $\hat{t}$ delineates the kinetic energy term, $\hat{v}_{\mathrm{loc}}$ and 
$\hat{v}_{\mathrm{nl}}$ correspond to the local and nonlocal pseudopotentials, respectively, 
$\hat{v}_{\mathrm{H}}$ represents the Hartree potential, and $\hat{v}_{\mathrm{XC}}$ denotes the 
exchange-correlation potential. Consequently, the one-body reduced density matrix is succinctly described by $\hat{\rho} = f(\hat{h}_{\mathrm{KS}}; \mu, \beta)$.

In sDFT, the electron density $\rho(\mathbf{r})$ is calculated using stochastic orbitals $|\chi\rangle$ as follows:
\begin{equation}
\rho(\mathbf{r}) = \left\langle\left\langle \chi \middle| \hat{\rho} \delta(\mathbf{r}-\hat{\mathbf{r}}) \middle| \chi \right\rangle\right\rangle_{\chi}
= \left\langle \left| \xi(\mathbf{r}) \right|^{2} \right\rangle_{\chi},
\label{eq:sdft-rho}
\end{equation}
where $\delta(\cdot)$ is the Dirac delta function, and $|\xi\rangle = 
\sqrt{\hat{\rho}}|\chi\rangle$ represents a projected stochastic orbital. 
The notation $\langle \cdots \rangle_{\chi}$ denotes averaging over all samples of $\chi$. In 
practical sDFT calculations, a finite number ($N_{\chi}$) of stochastic orbitals is employed. 
The stochastic orbital $\chi(\mathbf{r})$ is constructed to satisfy $\langle 
\chi(\mathbf{r})^{\ast} \chi(\mathbf{r}') \rangle_{\chi} = \delta(\mathbf{r}-\mathbf{r}')$. 
Practically, $\chi(\mathbf{r}) = \pm (\Delta V)^{-\frac{1}{2}}$, where 
$\Delta V = V/N_{\mathrm{G}}$ is the volume element of the real-space grid, and the sign of 
$\chi$ is randomly, uniformly, and independently selected for each grid point. Equation 
\ref{eq:sdft-rho} becomes exact in the limit as the number of stochastic orbitals $N_{\chi} \rightarrow \infty$; otherwise, the density obtained is a stochastic approximation.

The projection of stochastic orbitals onto the operator $\sqrt{\hat{\rho}}$ can be efficiently approximated by expanding $\sqrt{f(\hat{h}_{\mathrm{KS}};\mu,\beta)}$ through polynomial series, such as Chebyshev or Newton's Interpolation polynomials \cite{doi:10.1021/j100319a003,doi:10.1146/annurev.pc.45.100194.001045}. Specifically, this approximation takes the form:
\begin{equation}
\sqrt{f(\hat{h}_{\mathrm{KS}};\mu,\beta)} \approx \sum^{N_{\mathrm{c}}}_{n=0} a_{n}(\mu,\beta) T_{n}(\hat{h}_{\mathrm{KS}}),
\label{eq:cheby-expand}
\end{equation}
where $N_{\mathrm{c}}$ denotes the length of the polynomial chain, and $T_{n}(\cdot)$ are the Chebyshev polynomials. Within the framework of sDFT, the application of $T_{n}(\hat{h}_{\mathrm{KS}})$ to a stochastic orbital is computed utilizing the iterative formula associated with Chebyshev polynomials, thereby facilitating a computationally efficient projection algorithm.

A property $O$, other than electron density, can be calculated as follows. If $O$ is an explicit functional of 
electron density, it can be calculated using $\rho$ from Eq.~(\ref{eq:sdft-rho}). Otherwise, $O$ requires 
calculating the trace of an operator, which can be evaluated by the stochastic trace formula. In the stochastic 
trace formula, the trace of an operator $\hat{O}[\rho]$ is calculated with
\begin{equation}
    \mathrm{Tr}(\hat{O}[\rho]) = \langle\langle\chi|\hat{O}[\rho]|\chi\rangle\rangle_\chi\,.
    \label{eq:strace}
\end{equation}
Here, $\hat{O}[\rho]$ indicates that $\hat{O}$ depends on the electron density $\rho$. In a special case, 
$\hat{O}[\hat{\rho}] = \hat{\rho}[\rho]\hat{A}$, where $\hat{A}$ is a one-body operator. In this case,
\begin{equation}
    O = 2 \operatorname{Tr}(\hat{\rho} \hat{A}) = 2\langle\langle \xi | \hat{A} | \xi \rangle\rangle_{\chi}\,.
    \label{eq:operator}
\end{equation}
For some other properties, $\hat{O}$ is more complicated. For example, sDFT uses the following formula to 
calculate the electron density of states at energy $\varepsilon$: 
\begin{equation}
    D(\varepsilon) = \langle\langle\chi|\delta(\hat{h}_{\mathrm{KS}} - \varepsilon)|\chi\rangle\rangle_\chi
    \approx \langle\langle\chi|G(\hat{h}_{\mathrm{KS}}; \varepsilon, \sigma)|\chi\rangle\rangle_\chi,
    \label{eq:dos}
\end{equation}
where $G(x; \varepsilon, \sigma) = \frac{1}{\sqrt{2\pi\sigma^2}} e^{-(x-\varepsilon)^2 / 2\sigma^2}$ is a 
Gaussian function. A pre-selected $\sigma$ is the broadening of the Gaussian function, which determines the 
broadening of the density of states. The second example, electron entropy ($S_{\mathrm{KS}}$), is essential for 
finite temperature calculations involving metals. In sDFT, $S_{\mathrm{KS}}$ can be calculated using the 
formula:
\begin{equation}
    S_{\mathrm{KS}} = -2k_{\mathrm{B}} \langle\langle \chi | \hat{\rho} \log \hat{\rho} + (\hat{\mathrm{I}} - \hat{\rho}) \log (\hat{\mathrm{I}} - \hat{\rho}) | \chi \rangle\rangle_{\chi},
    \label{eq:entropy}
\end{equation}
where $\hat{\mathrm{I}}$ represents the identity operator. Both Eq.~(\ref{eq:dos}) and Eq.~(\ref{eq:entropy}) 
can be calculated with a Chebyshev polynomial expansion, similar to Eq.~(\ref{eq:cheby-expand}).

Due to the use of finite stochastic orbitals $|\chi\rangle$, noise exists in all quantities calculated with 
sDFT. This noise varies inversely with the square root of the number of stochastic orbitals, $N_{\chi}$, in 
accordance with the central limit theorem. Specifically, to reduce the stochastic fluctuation by one order of 
magnitude, it is necessary to increase $N_{\chi}$ by two orders of magnitude. Therefore, noise reduction 
techniques are crucial for accurately determining properties without compromising the efficiency of sDFT. 
It is important to note that for many observables, such as electron density, density of states, and atomic 
forces, the noise does not scale with the system size. This allows sDFT to maintain linear scaling. Conversely, 
for properties like the energy per particle, the noise diminishes as the system size increases. This implies 
that fewer stochastic orbitals are required for larger systems, making sDFT a sub-linear-scaling method. 
However, for total energy calculations, noise amplifies with system size, challenging the linear-scaling 
property of sDFT.
Users are encouraged to carefully assess the properties they intend to calculate with sDFT, considering the 
method's scalability and noise characteristics for their specific application.

\subsection{Overlapped Embedded Fragmentation}

Reduction in the statistical noise can be achieved by introduction of a reference system that can be calculated 
by deterministic KS-DFT. This will provide us a reasonable approximation of the electron density matrix which 
is corrected with sDFT. One way to generate such reference 
system is to decompose the whole system into overlapped 
fragments, leading to  the implementation of the overlapped embedded-fragmented stochastic DFT (o-efsDFT). We will 
outline the method in the rest of this section. 

\begin{figure}[H]
    \centering
    \includegraphics[scale=0.5]{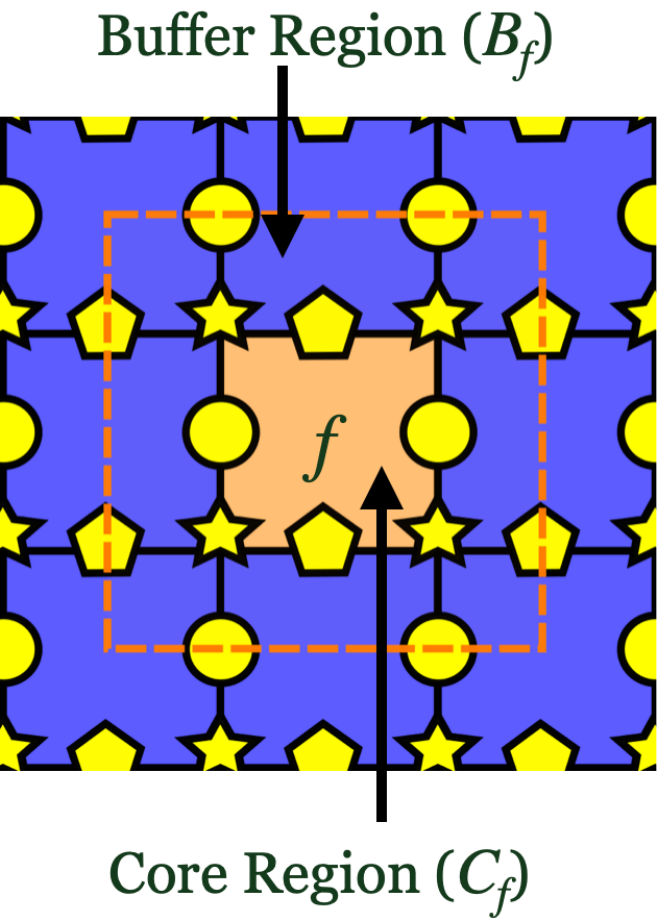}
    \caption{Schematic for overlapped fragmentation is shown in this figure. 
    \cite{artacho1999linear} The system is divided into overlapped fragments. 
    Each fragment $f$ is composed of a core region ($C_{f}$)
    and a buffer region ($B_{f}$)is defined around each $C_{f}$. A deterministic
    DFT is performed for each dressed fragment 
    ($D_{f}$), which is the union of $C_{f}$ and $B_{f}$ ($C_{f}\cup B_{f}$)}. 
    \label{fig:oefs-diagram}
\end{figure}

The o-efsDFT method divides a supercell into fragments refereed to as core regions ($C_{f}$) that is wrapped by a buffer region ($B_{f}$) to create a dressed fragment ($D_{f} = C_{f} \cup B_{f}$, where $f$ is the fragment index. The fragment density matrix, $\hat{\rho}_{f}$, is defined as 
\begin{equation}
\left\langle\mathbf{r}\middle|\hat{\rho}_{f}\middle| \mathbf{r}^{\prime}\right\rangle= \begin{cases}\sum_{i=1}\sqrt{f(\varepsilon_i^f;\mu_f,\beta)}\left\langle\mathbf{r} \middle| \varphi_{i}^{f}\right\rangle\left\langle\varphi_{i}^{f} \middle| \mathbf{r}^{\prime}\right\rangle, & \mathbf{r}' \in D_{f} \\ 0, & \mathbf{r}' \notin D_{f}\end{cases}
\label{eq:frag-dm}
\end{equation}
where $\varphi_{i}^{f}$ are the KS orbitals for the $f$'th fragment obtained from deterministic KS-DFT. $\varepsilon_i^f$ are fragment KS orbital energies. $\mu_f$ 
is the chemical potential that keeps each fragment 
charge neutral. We want to emphasize that using o-efsDFT 
to model a charge-separated system needs further developments. In o-efsDFT, electron density is calculated 
by
\begin{equation}
\rho(\mathbf{r}) =2 \sum_{f} \rho_{f}(\mathbf{r})+2\left\langle|\xi(\mathbf{r})|^{2}\right\rangle_{\chi}-2 \sum_{f}\left\langle\left|\xi_{f}(\mathbf{r})\right|^{2}\right\rangle_{\chi}\;\;,
\label{eq:oefsDFT-rho}
\end{equation}
where the fragment electron density is 
$\rho_{f}(\mathbf{r})=\sum_{i=1}f(\varepsilon_i^f,\mu_f,\beta)\left|\varphi_{i}^{f}(\mathbf{r})\right|^{2}$, $\xi_{f}(\mathbf{r})=\sum_{i=1}\sqrt{f(\varepsilon_i^f,\mu_f,\beta)} \varphi_{i}^{f}(\mathbf{r})\left\langle\varphi_{i}^{f} \middle| \chi\right\rangle_{D_{f}}$, and $\left\langle\varphi_{i}^{f} \middle| \chi\right\rangle_{D_{f}}=\int_{D_{f}} \mathrm{d} \mathbf{r} \varphi_{i}^{f}(\mathbf{r})^\ast \chi(\mathbf{r})$. 
If $\hat{\rho}_f$ is similiar with $\hat{\rho}$, noise 
in the second term of Eq.(\ref{eq:oefsDFT-rho}) will almost cancel noise in the third term. As we will demonsrate later, 
it is crucial to have a reasonably good fragment density 
matrix in order to maximizing the noise-redcution 
efficiency of sDFT. 

\subsection{Energy Windows}

Reduction in the stochastic noise can also be achieved by dividing the occupied space into subspaces named as ``energy windows''. In this method, instead of projecting the stochastic orbitals onto the occupied space, the stochastic orbitals are projected onto energy windows using a set of projectors, $\hat{\mathbf{P}}_{1}, \ldots, \hat{\mathbf{P}}_{N_{\mathrm{w}}}$. Here, the projector $\hat{\mathbf{P}}_{w}$ is defined as
\begin{equation}
\hat{\mathbf{P}}_{w}=f(\hat{h}_{\mathrm{KS}};e_w,\beta)-f(\hat{h}_{\mathrm{KS}};e_{w-1},\beta)
\end{equation}
for $1 \leq w \leq N_{\mathrm{w}}$, where $-\infty=e_{0}<e_1<\cdots<e_{N_{\mathrm{w}}}=\mu$. In energy windows sDFT (ew-sDFT), the electron density is obtained from the sum of all the projected densities. 
\begin{equation}
\rho(\mathbf{r})=2 \sum_{w=1}^{N_{\mathrm{w}}}\left\langle\left|\xi^{(w)}(\mathbf{r})\right|^{2}\right\rangle_{\chi} \equiv 2 \sum_{w=1}^{N_{\mathrm{w}}} \rho^{(w)}(\mathbf{r})\;\;,
\end{equation}
where $\left|\xi^{(w)}\right\rangle=\sqrt{\hat{\mathbf{P}}_{w}}|\chi\rangle$ is a projected stochastic orbital for window $w \cdot\left|\xi^{(w)}\right\rangle$ are calculated simultaneously with a single polynomial expansion. 

\subsection{Energy Window Embedded Fragmentation}

The final noise reduction method at our disposal combines the approaches used in o-efsDFT and ew-sDFT with the resulting expression for the electron density on each grid point $\textbf{r}$
\begin{align}
\rho(\mathbf{r}) = & 2 \sum_{f} \rho_{f}(\mathbf{r})+2 \sum_{w=1}^{N_{\mathrm{w}}}\left\langle\left|\zeta^{(w)}(\mathbf{r})\right|^{2}\right\rangle_{\chi} \nonumber \\
& -2 \sum_{f} \sum_{w=1}^{N_{\mathrm{w}}}\left\langle\left|\xi_{f}^{(w)}(\mathbf{r})\right|^{2}\right\rangle_{\chi} \;\;,
\end{align}
where $\zeta^{(w)}(\mathbf{r})=\left\langle\mathbf{r}\left|\sqrt{\hat{\rho} \hat{\mathbf{P}}_{w}}\right| \chi\right\rangle$ and 
$\xi_{f}^{(w)}(\mathbf{r})=\left\langle\mathbf{r}\left|\hat{\rho}_{f} \sqrt{\hat{\mathbf{P}}_{w}}\right| \chi\right\rangle$
The projection operators on the energy windows are 
the same as ew-efsDFT besides the last window is defined 
as 
\begin{equation}
\hat{\mathbf{P}}_{N_{\mathrm{w}}}=\hat{I}-\sum_{w=1}^{N_{\mathrm{w}}-1} \hat{\mathbf{P}}_{w},
\end{equation} 
In ew-sDFT, $\sum_{w=1}^{N_{\mathrm{w}}} \hat{\mathbf{P}}_{w}$ equals the density matrix $\hat{\rho}$, while in ew-efsDFT, it returns the identity operator, $\hat{I}$. In addition, the highest energy window is set to $\varepsilon_{N_{\mathrm{w}}}=\mu$ in ew-sDFT, while in ew-efsDFT, the energy windows are held fixed for the entire self-consistent procedure and are chosen to be independent of the chemical potential, $\mu$. This greatly simplifies the on-the-fly calculations of the chemical potential. \cite{10.1063/5.0044163}
The actions of $\sqrt{\hat{\rho} \hat{\mathbf{P}}_{w}}$ and $\sqrt{\hat{\mathbf{P}}_{w}}$ on $|\chi\rangle$ are obtained using a Chebyshev polynomial series. $\rho_f$ is calculated 
in the same way as o-efsDFT. 

\section{Results}

sDFT, o-efsFT, ew-sDFT, and ew-efsDFT were tested on an aluminum crystal of varying supercell sizes. Various temperatures ranging from 300 K to 60,000 K were used to test metal calculations at ambient conditions as well as in warm-dense metal systems. All calculations were performed using the Becke–Lee–Yang–Parr (BLYP) functional \cite{PhysRevA.38.3098, PhysRevB.37.785}.
It is important to emphasize that an exchange-correlation functional specifically tuned for warm-dense matter should be used to accurately model materials with extremely high electron temperatures \cite{doi:10.1021/acs.jpclett.2c03670, DORNHEIM20181, 10.1063/1.5143225}. However, this work mainly focuses on demonstrating the computational efficiency of various sDFT implementations; thus, the BLYP functional was used in all calculations. A wave function cutoff of 60.0 Ry and a density cutoff of 96.0 Ry, corresponding to a real-space grid spacing of 0.17 Å, were used for the calculations. A Troullier-Martins norm-conserving pseudopotential in Kleinman-Bylander form was employed \cite{troullier1991efficient, kleinman1982efficacious}. Throughout this study, 256 stochastic orbitals were used, regardless of the number of atoms, and 10 uniformly distributed energy windows were utilized in the ew-sDFT and ew-efsDFT calculations. All statistical quantities were obtained using 5 independent runs, unless explicitly stated otherwise.

\begin{figure}[H]
\centering
\includegraphics[scale=0.22]{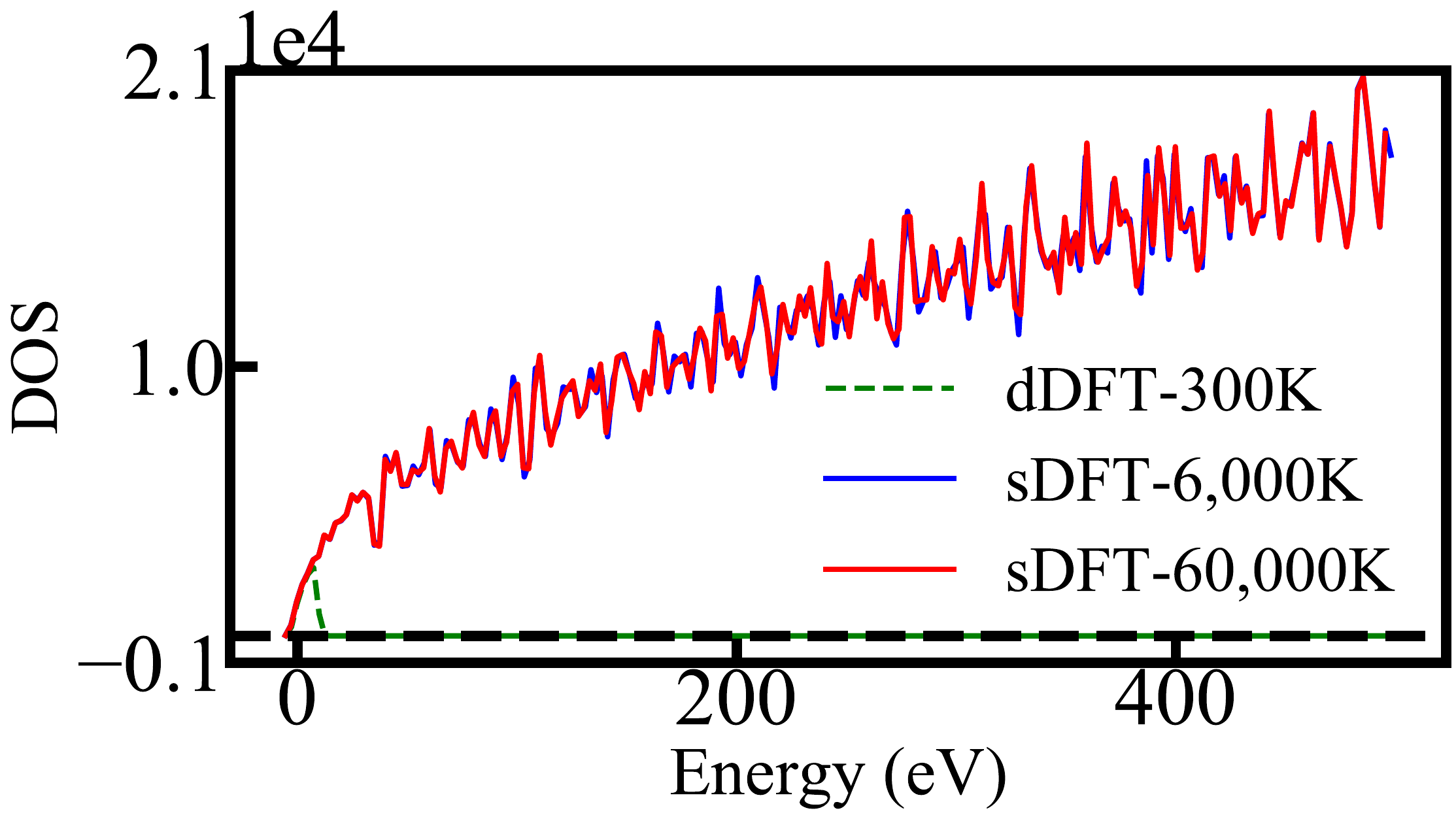}
\caption{\label{fig:epsart} The density of states of Al$_{256}$ at different electron temperatures are shown 
in this figure. The green dashed line is the density of states calculated from dDFT. The solid blue line 
and the solid red line are calculated from sDFT at temperature 6,000K and 60,000K with Eq.(\ref{eq:dos}).}
\label{dos}
\end{figure}

Previous work by White and Collins has demonstrated the advantages of sDFT in studying warm dense matter. \cite{white2020fast} The deterministic DFT (dDFT) calculations lack information on KS orbitals with high orbital energies. Therefore, evaluating properties which requires knowledge 
of high energy orbitals, like electron density of states, is not possbile with dDFT. In Fig.\ref{dos}, we demonstrate that sDFT is capable of estimating the electronic density of states even at extremely high energies. This is necessary for studying warm dense matter with high electron temperatures. Similar results were observed by White and Collins. \cite{white2020fast}

\begin{figure}[H]
\centering
\includegraphics[scale=0.18]{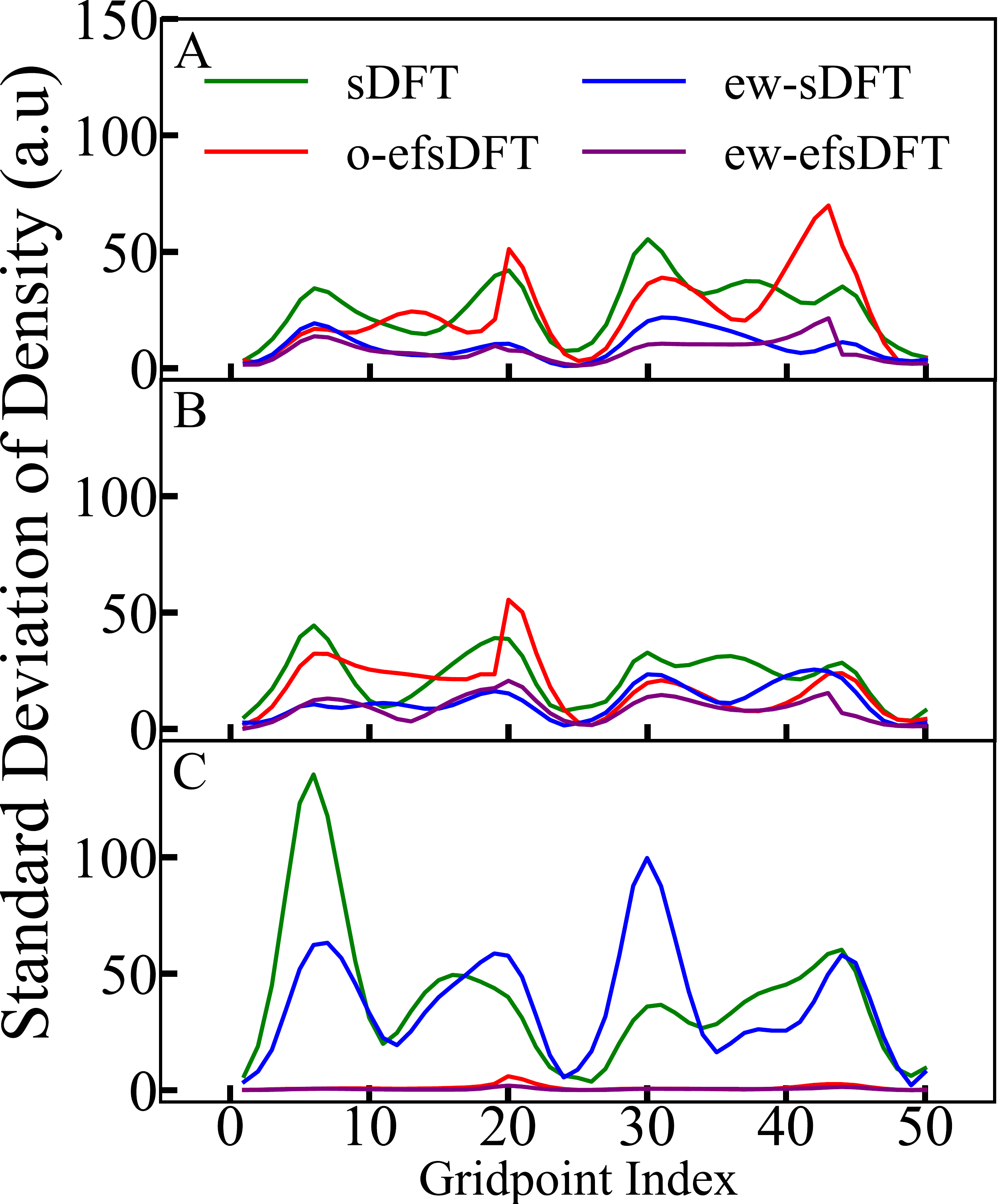}
\caption{\label{fig:density} This figure shows standard deviations (STD) of the density in sDFT and noise reduction techniques at different temperatures. We obtained the STDs over 5 independent trials for sDFT (green), ew-sDFT (blue), o-ef-sDFT (red), and ew-ef-sDFT (purple) at 300K (A), 6,000K (B), and 60,000K (C). We plotted the STDs for the first 50 grid points along the x-axis with $y = z = 0.0$ of Al$_{256}$.}
\label{std-density}
\end{figure}

Fig.\ref{std-density} shows the standard deviations (STD) of the electron densities calculated with various sDFT implementations at different electron temperatures. Panel A demonstrates that real-space fragmentation is not capable of reducing noise in electron density at low electron temperatures; specifically, the noise from o-efsDFT is comparable to that from sDFT, and the noise from ew-efsDFT is comparable to that from ew-sDFT. On the other hand, the energy window method works reasonably well and can reduce the noise in electron density by almost 50$\%$. This qualitative noise reduction performance is similar when the electron temperature is increased to 6000 K, as demonstrated in Fig.\ref{std-density}, Panel B. Since the Fermi-Dirac function becomes smoother at higher electron temperatures, the overlap between two neighboring energy windows becomes larger. Therefore, both ew-sDFT and ew-efsDFT become less effective at reducing noise. Real-space fragmentation still cannot efficiently reduce noise in electron density at 6000 K. When the electron temperature is increased to 60,000 K, the efficiency of noise reduction becomes very different, as shown in Panel C of Fig.\ref{std-density}. First, all real-space-fragmentation-based methods, including o-efsDFT and ew-efsDFT, can significantly reduce noise in electron density. The efficiency changes for real-space fragmentation methods will be explained in the following paragraph. Second, using only ew-sDFT is not capable of reducing noise in the electron density because there is a significant overlap between two neighboring window functions.

\begin{figure}[H]
\centering
\includegraphics[scale=0.2]{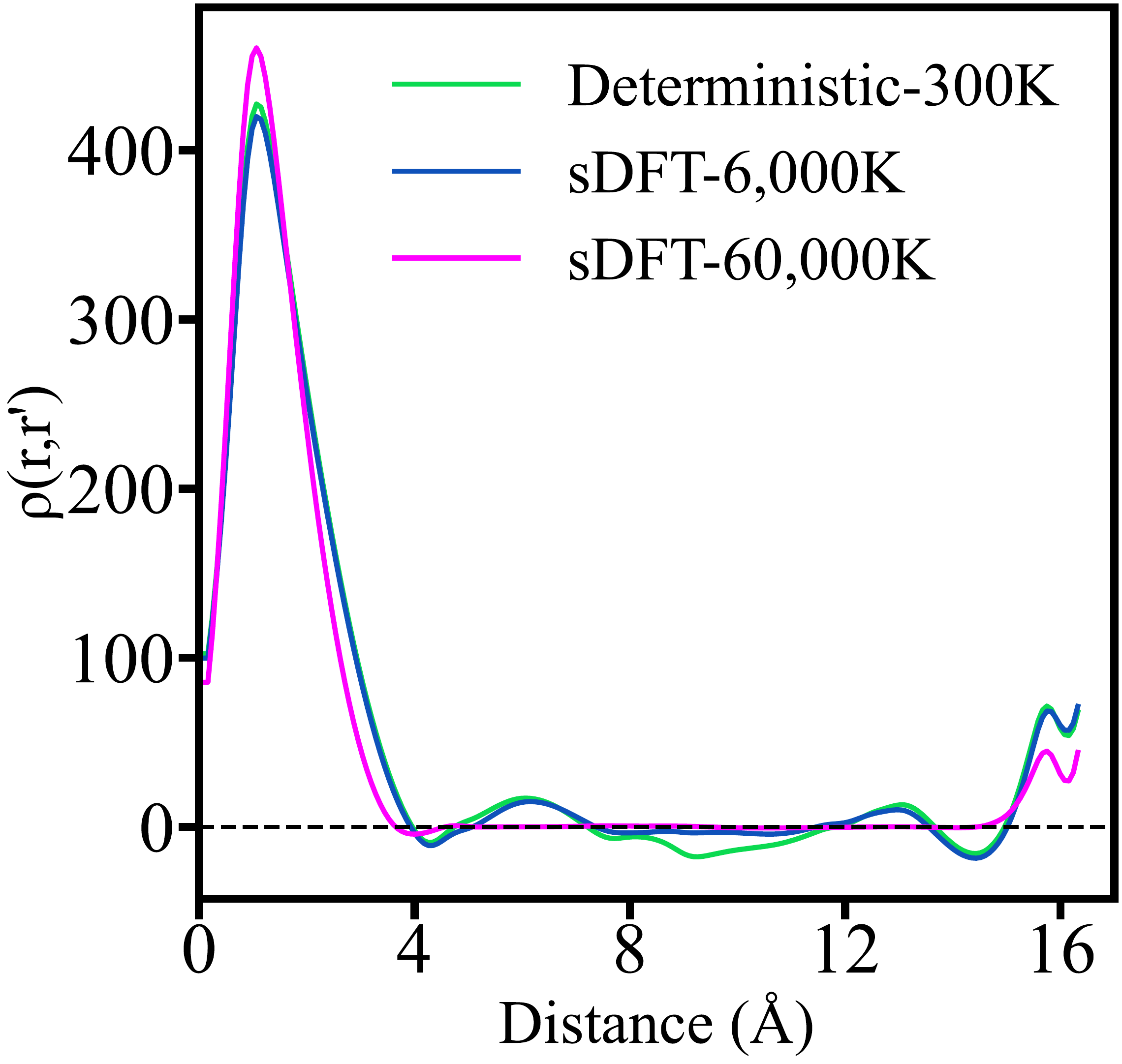}
\caption{\label{fig:epsart} One-dimensional slices of density matrix is shown in this figure. 
The green line is the results from a 300K deterministic DFT calculation while the blue/magenta line is from a 
sDFT calculation at 6000 K/60000 K. Details of calculating the density matrix are presented in the 
support information. }
\label{density-matrix}
\end{figure}

It has been demonstrated that the noise reduction efficiency of real-space fragmentation methods depends on the difference between the fragment density matrix $\sum_f\hat{\rho}_f$ and the system density matrix $\hat{\rho}$. \cite{10.1063/1.5064472, 10.1063/5.0044163} The difference between $\sum_f\hat{\rho}_f$ and $\hat{\rho}$ arises for two reasons. First, a $\Gamma$-point calculation is performed for each fragment. The finite size error from using a small supercell in a fragment DFT calculation is one reason for the discrepancy between $\sum_f\hat{\rho}_f$ and $\hat{\rho}$. Second, the fragment density matrix $\hat{\rho}_f$ is truncated at the boundary of $D_f$, according to Eq.(\ref{eq:frag-dm}). At 300 K, $\hat{\rho}_f$ is very sensitive to fragment size, which can be confirmed by performing a unit cell calculation with k-point sampling. At this temperature, DFT calculations converge slowly with an increasing number of k-points (see Fig.S1 in the Supplemental Information). Also, $\hat{\rho}_f$ is highly delocalized at 300 K, as shown in Fig.\ref{density-matrix}, leading to a large truncation error at the boundary of $D_f$. However, at 60,000 K, fragment DFT calculations converge quickly with an increasing number of k-points, and $\hat{\rho}_f$ decays to zero much faster 
(see Fig.\ref{density-matrix}). Therefore, using a small supercell in a fragment DFT calculation results in $\sum_f\hat{\rho}_f$ that is similar to $\hat{\rho}$. Errors associated with truncating $\hat{\rho}_f$ are also negligible at 60,000 K.

\begin{table*}
\caption{\label{energies} The average energies per electron and corresponding standard deviations (STD) for various sDFT methods at different temperatures. The statistics were obtained from 5 independent trials for each method of sDFT used in this study. In order, the kinetic energy ($E_{\mathrm{K}}$), non-local pseudopotential energy ($E_{\mathrm{nl}}$), local pseudopotential energy ($E_{\mathrm{loc}}$), hartree energy ($E_{\mathrm{H}}$), exchange-correlation energy ($E_{\mathrm{XC}}$), entropy ($TS_{\mathrm{KS}}$), and total energy ($E_{\mathrm{tot}}$) are shown in units of eV in the table. Each entry represents the average energy up to 4 decimal places and in parenthesis next to it, is the STD on the order of 0.1 meV.}
\begin{ruledtabular}
\begin{tabular}{ccccccccc}
 Temperature(K)&Method&$E_{\mathrm{K}}$&$E_{\mathrm{nl}}$
 &$E_{\mathrm{loc}}$&$E_{\mathrm{H}}$&$E_{\mathrm{XC}}$&$TS_{\mathrm{KS}}$&$E_{\mathrm{tot}}$\\ \hline
 
300&Deterministic&7.7052&-0.3639&5.4858&0.0381&-6.8521&-0.0011&-18.7095 \\
 &sDFT&7.7215(154)&-0.3666(12)&5.4858(16)&0.0494(3)&-6.8582(7)&-0.0009(0)&-18.6905(140) \\
 &ew-sDFT&7.7059(124)&-0.3645(11)&5.4863(5)&0.0397(2)&-6.8531(2)&-0.0011(0)&-18.7084(110) \\
 &o-efsDFT&7.7063(26)&-0.3643(3)&5.4871(13)&0.0434(2)&-6.8598(4)&-0.0010(0)&-18.7098(18) \\
 &ew-efsDFT&7.7097(27)&-0.3643(5)&5.4861(3)&0.0385(2)&-6.8528(2)&-0.0011(0)&-18.7054(24) \\
 
 &&&&&&\\
 
6,000&Deterministic&7.7702&-0.3733&5.4846&0.0395&-6.8534&-0.1193&-18.7731 \\
&sDFT&7.8035(250)&-0.3763(14)&5.4854(12)&0.0505(2)&-6.8595(3)&-0.1198(7)&-18.7377(227) \\
 &ew-sDFT&7.7842(205)&-0.3747(16)&5.4853(8)&0.0431(2)&-6.8556(3)&-0.1196(8)&-18.7587(191) \\
 &o-efsDFT&7.7696(45)&-0.3735(6)&5.4852(9)&0.0424(3)&-6.8576(4)&-0.1194(10)&-18.7748(40) \\
 &ew-efsDFT&7.7739(24)&-0.3737(3)&5.4846(3)&0.0401(1)&-6.8543(1)&-0.1199(11)&-18.7706(15) \\

&&&&&&\\
 
 20,000&Deterministic&8.4300&-0.4342&5.4795&0.0368&-6.8528&-1.3115&-19.3737 \\
&sDFT&8.4596(196)&-0.4372(16)&5.4803(10)&0.0482(4)&-6.8599(4)&-1.3159(30)&-19.3464(193) \\
 &ew-sDFT&8.4677(230)&-0.4375(17)&5.4798(10)&0.0452(6)&-6.8582(5)&-1.3161(36)&-19.3407(200) \\
 &o-efsDFT&8.4311(22)&-0.4344(4)&5.4797(2)&0.0373(1)&-6.8538(1)&-1.3144(28)&-19.3760(32) \\
 &ew-efsDFT&8.4300(37)&-0.4342(3)&5.4797(1)&0.0370(0)&-6.8532(1)&-1.3145(38)&-19.3766(52) \\

\end{tabular}
\end{ruledtabular}
\label{tab: energy}
\end{table*}

Noise reduction techniques can also be applied to other ground state properties like ground state energy and atomic forces. We list different energy terms calculated with various methods at different temperatures in Table \ref{tab: energy}. We would like to emphasize that the reason we display results only up to 20,000 K is due in part to deterministic calculations of Al$_{256}$ at a temperatures higher than 20,000 K is highly challenging. For all ground state energies, including those for sDFT above 20,000 K, please refer to Table S1 in the Supplemental Information. Some energy terms, such as local pseudopotential energy, Hartree energy, and exchange-correlation energy, can be directly evaluated with electron density. The noise in these energy terms is consistent with the noise in electron density. At 300 K, the noise reduction efficiency of o-efsDFT is limited, while o-efsDFT can significantly reduce noise at 20,000 K. ew-sDFT works well at 300 K, and its noise reduction efficiency is dramatically reduced at 20,000 K. For $E_{\mathrm{loc}}$, $E_{\mathrm{H}}$, and $E_{\mathrm{XC}}$, ew-efsDFT performs best at all three temperatures.

Kinetic energy and non-local pseudopotential energy can be calculated using Eq.~(\ref{eq:operator}). 
Noise reduction methods for these energy terms have been developed in previous 
studies~\cite{10.1063/1.5064472, 10.1063/1.5114984, 10.1063/5.0044163}. 
Real-space-fragmentation-based methods, such as o-efsDFT and ew-efsDFT, 
have demonstrated significant noise reduction in these energy terms. However, ew-sDFT 
can only achieve marginal noise reduction, even at low electron temperatures. 
Previous studies have indicated that, 
at low electron temperatures, ew-sDFT can not reduce the noise in $\mathrm{Tr}(\hat{\rho}\hat{O})$ only if the matrix representation 
of $\hat{O}$ is diagonal-dominant in the deterministic Kohn-Sham (KS) orbital basis~\cite{10.1063/1.5114984}. 
Our results suggest that the kinetic operator $\hat{t}$ and the non-local pseudopotential operator $\hat{v}_{\mathrm{nl}}$ 
tend to be diagonal-dominant in the deterministic KS orbital basis for the tested system. 

Noise in the electron entropy term, $S_{\mathrm{KS}}$, originates from two primary sources. 
The first source is the noise in electron density, which leads to noise in the Kohn-Sham Hamiltonian, $\hat{h}_{\mathrm{KS}}$, and subsequently in the density matrix, $\hat{\rho}$. 
The second source is inherent in the stochastic trace formula used to calculate entropy. 
Despite significant reductions in electron density noise through methods such as o-efsDFT, ew-sDFT, 
and ew-efsDFT, we observe that noise in $TS_{\mathrm{KS}}$ persists. This observation indicates that 
the dominant noise contribution arises from the stochastic trace formula rather than electron density 
fluctuations.
To delve deeper into this issue, we utilized converged electron densities, $\rho_{\mathrm{sDFT}}$, from sDFT or 
its noise-reduced variants, and performed a deterministic diagonalization of 
$\hat{h}_{\mathrm{KS}}[\rho_{\mathrm{sDFT}}]$ to evaluate Eq.~(\ref{eq:entropy}). 
This deterministic trace approach revealed a decrease in entropy noise when employing noise reduction 
techniques (see Table S2 in the Supplemental Information). This outcome underscores that the primary 
source of entropy noise is the stochastic trace formula, overshadowing the impact of noise reduction 
in electron density. Addressing the noise in the stochastic trace formula for entropy thus necessitates 
further research and development.
Nevertheless, our findings indicate that the presence of significant noise in the entropy term does not 
adversely affect the accuracy of calculated electron densities and atomic forces. 

In addition to noise-related challenges, we also identified the presence of bias or systematic errors in 
certain energy terms, particularly those that are non-linear functionals of electron 
density~\cite{https://doi.org/10.1002/wcms.1412}. Notably, bias is evident in 
$E_{\mathrm{H}}$, $E_{\mathrm{XC}}$, and $TS_{\mathrm{KS}}$. Such biases are inherent to sDFT calculations, 
where, for instance, $E_{\mathrm{H}}$ obtained from sDFT is significantly higher compared to dDFT calculations, despite relatively small fluctuations in $E_{\mathrm{H}}$.
Noise-reduction methods play a crucial role in mitigating these biases, largely due to their effectiveness in minimizing fluctuations in electron density. Specifically, the systematic error in all energy terms is notably 
reduced when employing ew-efsDFT.
\begin{figure}[H]
\centering
\includegraphics[scale=0.185]{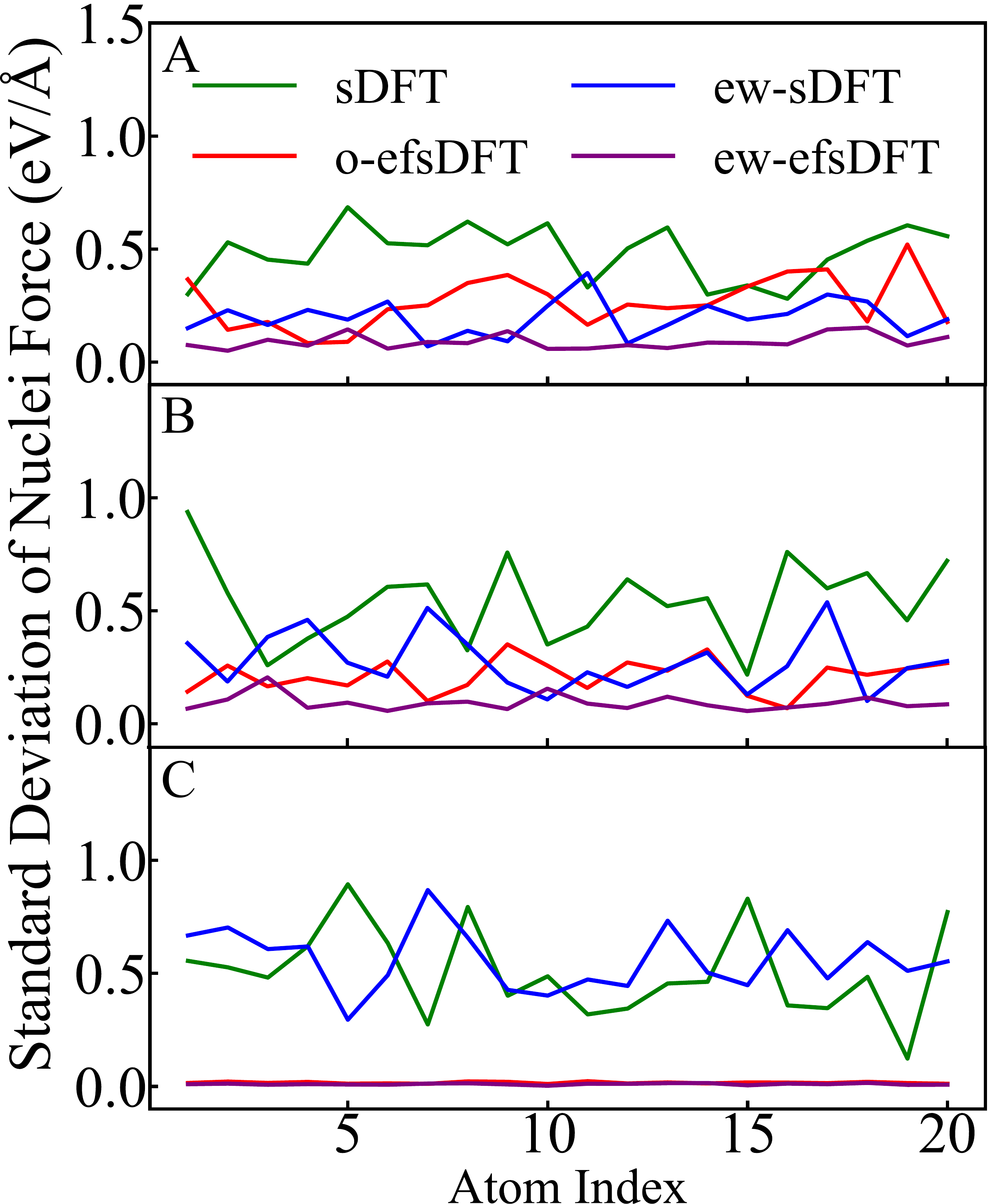}
\caption{\label{fig:epsart} Standard deviations (STD) of the total nuclei force (eV/Å)in sDFT and noise reduction techniques at different electron temperatures. We obtained the STD over 5 independent trials for sDFT (green), ew-sDFT (blue), o-ef-sDFT (red), and ew-ef-sDFT (purple) at 300 K (A), 6,000 K (B), and 60,000 K (C). We plotted the STDs for the first 20 atoms of Al$_{256}$.}
\label{std-force}
\end{figure}

\begin{figure}[H]
\centering
\includegraphics[scale=0.17]{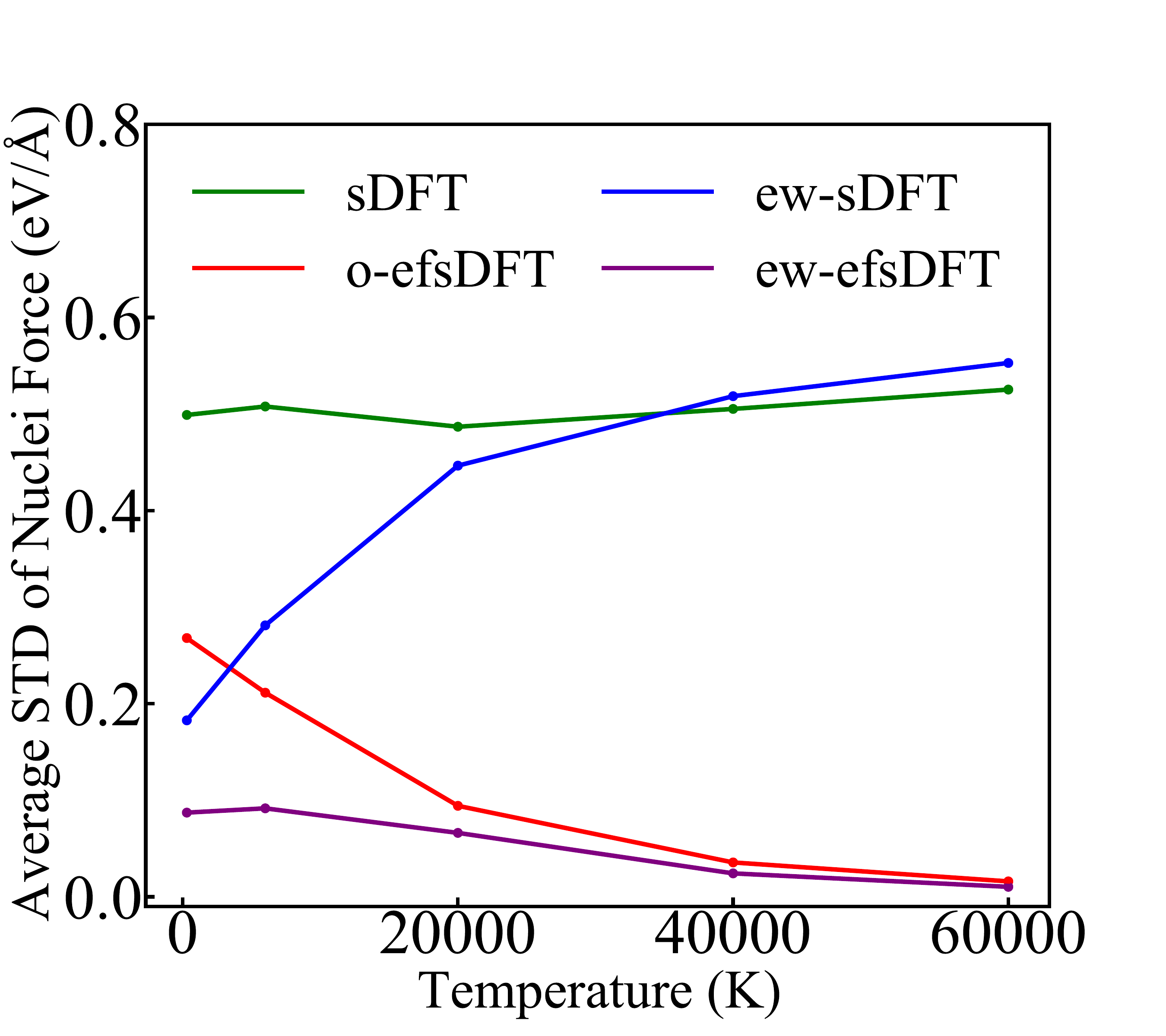}
\caption{\label{fig:epsart} Noise reduction efficiency in nuclei forces (eV/Å) of various sDFT methods with respect to temperature. 
We obtained the standard deviation (STD) of atomic force on each atom over 5 independent trials for each method at 300 K, 6000 K, 20,000 K, 40,000 K, and 60,000 K. We then average all STDs of different atoms and plotted the corresponding values for sDFT (green), ew-sDFT (blue), o-ef-sDFT (red), and ew-ef-sDFT (purple).}
\label{std-force-trend}
\end{figure}

Reducing noise in atomic forces is very important for determining equilibrium structures.\cite{arnon2020efficient, chen2023structure} In Fig.\ref{std-force}, the STD of the force on the nuclei along the x-axis ($F_{x}$) for the first 20 atoms is plotted for sDFT and noise reduction methods at electron temperatures of 300 K (Panel A), 6,000 K (Panel B), and 60,000 K (Panel C). At 300K, both o-efsDFT and ew-sDFT can marginally improve the fluctuation of nuclear forces, while ew-efsDFT performs best at 300 K. At 60,000 K, ew-sDFT is not capable of reducing noise in nuclear forces. However, both methods, o-efsDFT and ew-efsDFT, can significantly reduce the noise in nuclear forces at 60,000 K. This is consistent with the noise reduction efficiency observed in electron density at different temperatures. In Fig.\ref{std-force-trend}, we highlight the efficiency of sDFT and noise reduction techniques by plotting the average STD of nuclear forces along the x-axis as a function of temperature. The efficiency of ew-sDFT is consistent with the work done by Hadad et al.\cite{hadad2024stochastic}.

To investigate the impact of system size on the noise reduction efficiency of sDFT and ew-efsDFT, we conducted a series of tests focusing on the atomic forces within different-sized systems. Our tests involved analyzing data from five independent trials for both sDFT and ew-efsDFT across a range of system sizes. However, due to resource constraints, we were limited to single runs for Al$_{500}$ and Al$_{864}$ when using

\noindent  ew-efsDFT. Given these limitations, it was crucial to validate the reliability of our statistical analysis on the atomic forces for these larger systems. To achieve this, we employed two distinct statistical methods for smaller system sizes, which served as a basis for comparison. The first method calculated the STD of the force on each atom across the independent trials, providing a measure of variability for individual atomic forces. The second method aggregated the forces from all atoms within a single trial and then calculated the STD, offering a holistic view of force variability within a single system snapshot. Our findings, presented in Table \ref{tab:force-statstics-system-size-6000k}, reveal no significant difference between the results obtained from these two statistical approaches. This consistency suggests that, for larger systems where conducting multiple calculations is not feasible due to resource constraints, a single calculation can reliably represent the statistical behavior of atomic forces. This insight is particularly valuable for efficiently assessing the noise reduction efficiency of sDFT and ew-efsDFT in large-scale systems, indicating that even with limited data, meaningful statistical conclusions can be drawn. We want to emphasize that the equivalence of the two statistical methods may only be applicable to a crystal structure, where atoms are equivalent in such a system. We do not recommend calculating STD across different atoms in a slab model or a nanocrystal. Aside from the sDFT Al$_{4}$ results in Table \ref{tab:force-statstics-system-size-6000k}, which could be the result of artifacts from using a single unit cell, there appear to be no significant changes in the efficiency of noise reduction in response to system size. Since the number of stochastic orbitals is kept constant across all calculations, this suggests that the number of stochastic orbitals needed for a system is nearly independent of the system size.

\begin{table}[H]
\caption{\label{sdft-ewefs-force-stats} Comparison of different methods of obtaining statistics for nuclei forces (eV/Å). Here we compare using sDFT and ewef-sDFT on various system sizes of Al all of which were ran at an electron temperature of 6,000K. The Numbers reported in column A was a result of taking the STD across all atoms for each trial, then taking the average of those STDs. Numbers reported in column B was a result of taking the STD for each atom across all trials, then taking the average of those STDs. For Al$_{500}$ and Al$_{864}$, statistics were obtained from only B.}
\begin{ruledtabular}
\begin{tabular}{lcccc}
 Temperature(K)&Method&System&A&B\\ \hline
 6,000&sDFT&Al$_{4}$&0.3258&0.3332\\
 &&Al$_{32}$&0.6225&0.5257\\
 &&Al$_{108}$&0.5605&0.4764\\
 &&Al$_{256}$&0.6051&0.5078\\
 &&&&\\
 &ew-efsDFT&Al$_{108}$&0.1193&0.1024\\
 &&Al$_{256}$&0.1078&0.0917\\
 &&Al$_{500}$&------&0.1097\\
 &&Al$_{864}$&------&0.1146\\
\\
60,000&sDFT&Al$_{4}$&0.3434&0.3476\\
 &&Al$_{32}$&0.6756&0.5861\\
 &&Al$_{108}$&0.5939&0.5079\\
 &&Al$_{256}$&0.6222&0.5253\\
 &&&&\\
 &ew-efsDFT&Al$_{108}$&0.0165&0.0140\\
 &&Al$_{256}$&0.0122&0.0104\\
 &&Al$_{500}$&------&0.0154\\
 &&Al$_{864}$&------&0.0170\\
\end{tabular}
\end{ruledtabular}
\label{tab:force-statstics-system-size-6000k}
\end{table}

\begin{figure}[H]
\centering
\includegraphics[scale=0.19]{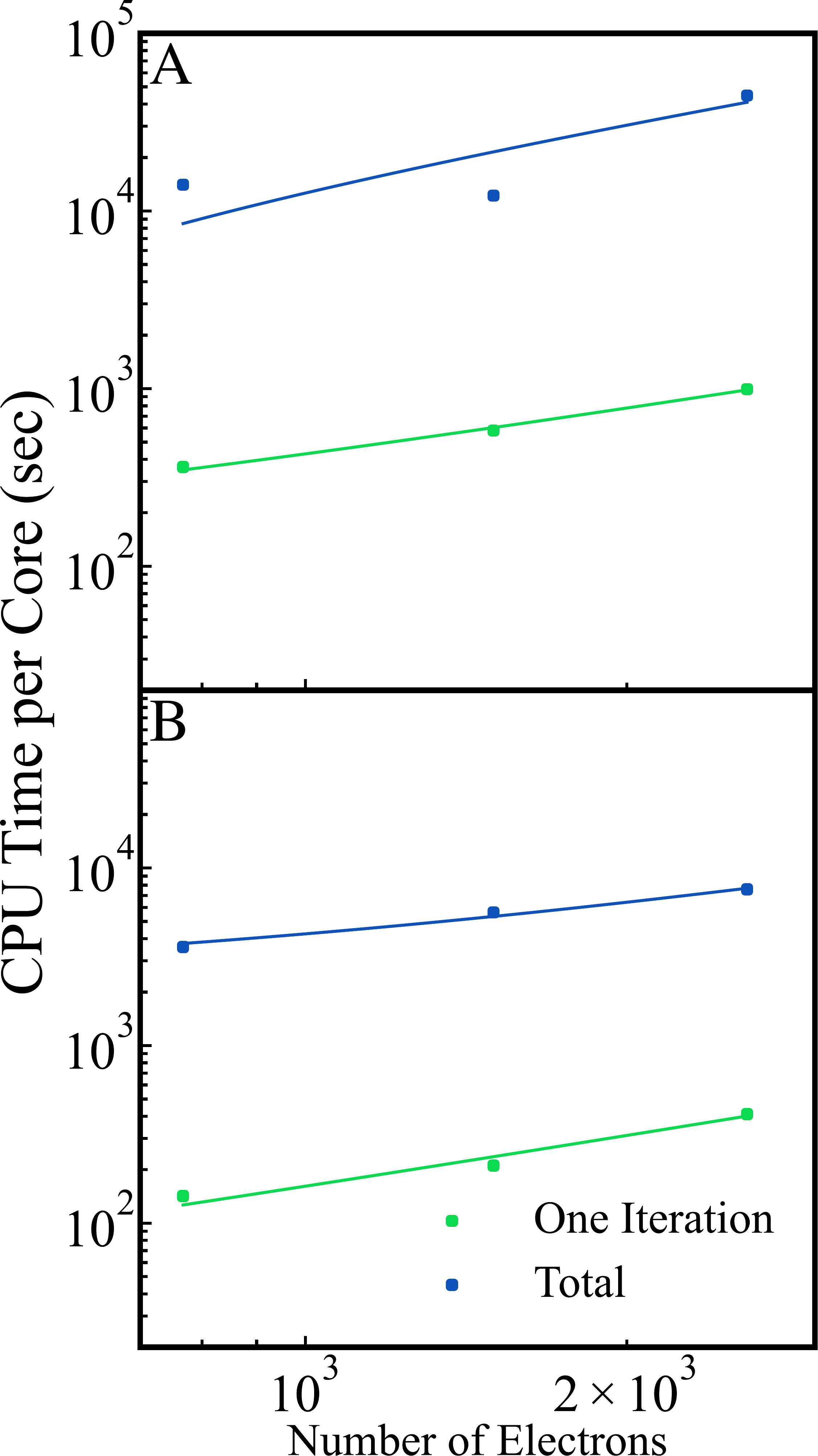}
\caption{\label{fig:epsart} Scaling of ew-efsDFT at 900 K and 6,000 K on NERSC Corri-KNL. In panels A (6,000 K) and B (900 K), the average time it took to complete one SCF iteration (blue) and the total time (green) to converge the SCF calculations were plotted against their corresponding system size. }
\label{scaling-knl}
\end{figure}

\begin{figure}[H]
\centering
\includegraphics[scale=0.19]{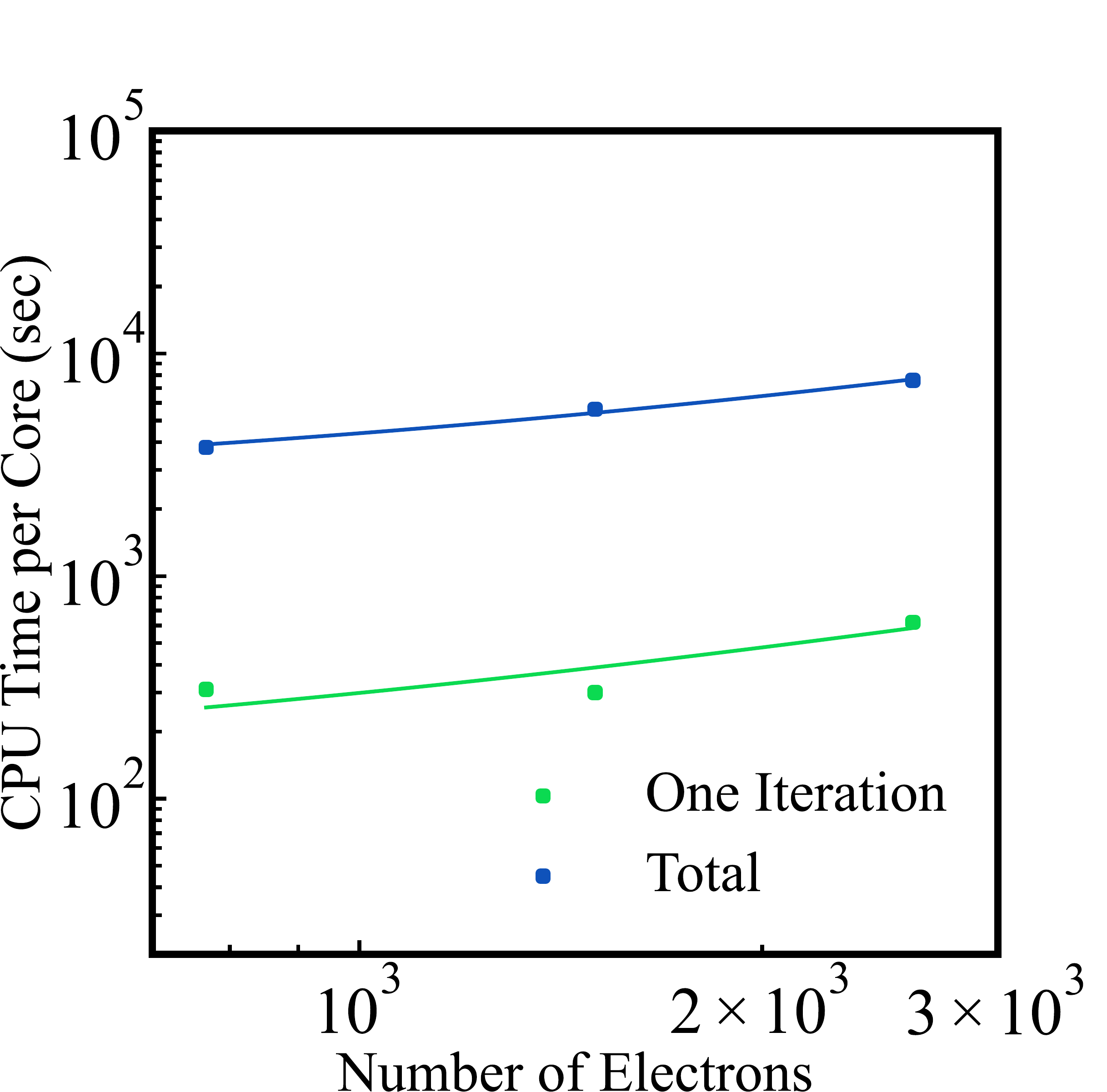}
\caption{\label{fig:epsart} Scaling of ew-efsDFT at 60,000 K on NERSC Perlmutter. The average time it took to complete one SCF iteration (blue) and the total time (green) to converge the SCF calculations were plotted against their corresponding system size.}
\label{scaling-perl}
\end{figure}

Given the performance of ew-efsDFT in noise reduction for properties such as density, ground-state energy, and force on nuclei, we tested its computational efficiency for larger systems. Our tests were conducted on the Cori-KNL and Perlmutter supercomputers at the National Energy Research Scientific Computing Center (NERSC). Calculations on Cori-KNL included Al$_{256}$, Al$_{500}$, and Al$_{864}$ at 900 K and 6,000 K. The same set of system sizes was tested on Perlmutter, but at a system temperature of 60,000 K. The system parameters defined earlier in the discussion were applied to all six runs. In Fig.\ref{scaling-knl} and Fig.\ref{scaling-perl}, we plotted the total and per iteration CPU time per core versus the number of electrons in the system on a log-log scale. The top panel of Fig.\ref{scaling-knl} presents the results for the runs at 6,000 K, which shows that the time needed for each iteration scales as $O(N{e}^{0.83})$, while the time needed for a single calculation scales as $O(N_{e}^{0.91})$. The bottom panel of Fig.\ref{scaling-knl} displays the results for the runs at 900 K, indicating that the time needed for each iteration scales as $O(N_{e}^{0.86})$, while the time needed for a single calculation scales as $O(N_{e}^{0.62})$. In Fig.\ref{scaling-perl}, the run conducted on Perlmutter at 60,000 K showed that the time needed for each iteration scales as $O(N_{e}^{0.63})$, and the time needed for a single calculation scales as $O(N_{e}^{0.96})$. We continue to observe similar scaling on different computational architectures, and the scaling does not seem to be impacted by system size or electron temperature.

Fig.(\ref{scaling-perl}) suggests that our noise reduction method, ew-efsDFT, is significantly different from the mixed stochastic density functional theory (MDFT) developed by White and Collins. \cite{white2020fast} In MDFT, stochastic orbitals are used to correct a dDFT calculation at a lower electron temperature. Therefore, the scaling of MDFT is still limited by the need to perform a dDFT calculation for the whole system. In contrast, ew-efsDFT uses stochastic orbitals to correct a reference system composed of fragments. By taking advantage of the localized density matrix at high electron temperatures, the noise reduction efficiency of ew-efsDFT is significant, and linear scaling is still achieved at high electron temperatures. This enables the modeling of disordered warm dense matter with a large supercell.

\section{Conclusions}
In this work, we present benchmark studies on noise-reduction techniques in stochastic 
density functional theory (sDFT) for metallic systems. Our findings indicate that the 
efficiency of noise reduction depends on the electron temperature. Specifically, 
at low electron temperatures, the energy-window-based method, ew-sDFT, outperforms the 
fragmentation-based approach, o-efsDFT. Conversely, at extremely high electron 
temperatures, o-efsDFT demonstrates superior performance over ew-efsDFT, 
attributed to the localized nature of the one-body density matrix. At both low and high 
electron temperatures, ew-efsDFT, which integrates both the energy-window and 
fragmentation approaches, emerges as the most efficient method.
Our results further reveal that ew-efsDFT scales linearly, enabling the efficient 
simulation of large metallic systems across a large range of electron temperatures. 
Despite 
ew-efsDFT's current status as the most effective noise-reduction method, we observe 
significant residual noises in electron density, energy, and nuclear forces at low electron temperatures. This underscores the necessity for further advancements. Additionally, the development of a formula to mitigate noise in electron entropy is imperative to enhance the accuracy of total energy calculations at high electron temperatures.

\section*{DATA AVAILABILITY}
The data that support the findings of this study are available from the corresponding author upon reasonable request.

\section*{Supplementary Material}
Discussions on convergence of electronic structure with system size and electron temperatures are 
included in the Supplementary Material \cite{Giannozzi2009,Giannozzi2017}. 
The Supplementary Material also includes a complete 
data on energies per electron with different electron temperatures, analysis of noise in entropy, 
and details in calculating the density matrix. 

\section{Acknowledgments}
J. V. and M. C. gratefully acknowledges support from the National Science Foundation EAR-2246687. 
Resources of the National Energy Research Scientific Computing Center (NERSC), a U.S. Department of Energy Office of Science User Facility operated under Contract No. DE-AC02-05CH11231, are greatly acknowledged.

\providecommand{\noopsort}[1]{}\providecommand{\singleletter}[1]{#1}%

\end{document}